\documentclass[12pt]{article}

\newcommand{\be}{\begin{equation}}
\newcommand{\ee}{\end{equation}}
\newcommand{\bea}{\begin{eqnarray}}
\newcommand{\eea}{\end{eqnarray}}

\font\mybb=msbm10 at 12pt

\def\bb#1{\hbox{\mybb#1}}

\def\Re {\bb{R}}

\def\id{\protect{{1 \kern-.28em {\rm l}}}}

\makeatletter
\renewcommand\section{\@startsection {section}{1}{\z@}%
                                   {-3.5ex \@plus -1ex \@minus -.2ex}%
                                   {2.3ex \@plus.2ex}%
                                   {\normalfont\large\bfseries}}
\renewcommand\subsection{\@startsection{subsection}{2}{\z@}%
                                   {-3.25ex\@plus -1ex \@minus -.2ex}%
                                   {1.5ex \@plus .2ex}%
                                   {\normalfont\normalsize\bfseries}}
\makeatother

\begin{document}

\begin{titlepage}

$~$
\bigskip\bigskip
\bigskip\bigskip

\centerline{\Large On spin chains and field theories}

\bigskip
\bigskip
\centerline{  Radu Roiban}
\bigskip
\medskip

\bigskip
\centerline{\it Department of Physics}
\centerline{\it University of California, Santa Barbara, CA\ \ 93106}
\centerline{ radu@vulcan2.physics.ucsb.edu}
\bigskip
\bigskip

\begin{abstract}

We point out that the existence of global symmetries in a field theory
is not an essential ingredient in its relation with an integrable
model. We describe an obvious construction which, given an integrable
spin chain, yields a field theory whose 1-loop scale transformations 
are generated by the spin chain Hamiltonian. We also identify
a necessary condition for a given field theory to be related to an
integrable spin chain.

As an example, we describe an anisotropic and parity-breaking
generalization of the XXZ Heisenberg spin chain and its associated
field theory. The system has no nonabelian global symmetries and 
generally does not admit a supersymmetric extension without the
introduction  
of more propagating bosonic fields. For the case of a 2-state chain
we find the spectrum and the eigenstates.
For certain values of its coupling constants the field theory
associated to this general type of chain is the bosonic sector of the
q-deformation of ${\cal N}=4$ SYM theory. 

\end{abstract}

\end{titlepage}

\section{Introduction}

The dilatation operator at 1-loop level in ${\cal N}=4$ supersymmetric
Yang-Mills theory (SYM) has
been  subject to extensive investigation over the past few years. A
surprising result was that its action on gauge invariant operators
can be mapped into the action of the Hamiltonian of a spin chain on
the vectors in its Hilbert space. The first steps in this direction
were taken in  \cite{MIZA} where it was shown that the action of the 
1-loop dilatation operator on scalar operators can be described
by an $SO(6)$ spin chain in the fundamental representation. 
Other bosonic sectors of ${\cal N}=4$ SYM enjoy similar properties (see
e.g. \cite{BELI_N4}). Combining these results as well as some partial
supersymmetric extensions \cite{BEI1} led to the
realization that the dilatation operator for the full theory is
described by an $PSU(2,2|4)$ spin chain \cite{BEI_full}. In this
context, the continuum limit of some of the bosonic spin chains and
their relation to a two-dimensional sigma model were discussed in
\cite{KRUZ}. 

An apparently different integrable structure was found in the
world sheet theory of the bosonic string \cite{msw} and superstring
\cite{BEPR} in $AdS_5\times S^5$. 
Through the AdS/CFT correspondence, this translates into a certain
algebraic structure 
in ${\cal N}=4$ SYM which was shown \cite{DONW} to be
consistent with the $PSU(2,2|4)$ spin chain representation of the
dilatation operator.

More involved collections of operators in ${\cal N}=4$ SYM also have 
certain relations with integrable models. The world sheet theory of
strings in $AdS_5\times S^5$  expanded around certain semiclassical
solutions \cite{spinning} has a description in terms of integrable
models (see \cite{sp_int} and \cite{TSEY_rev} for a review). 
Then, the
AdS/CFT correspondence implies that the set of operators dual to
excitations around this semiclassical configurations should exhibit a
similar integrable structure.

The dilatation operator at higher loops can also be described as 
an operator acting on the spin chain appearing at the 1-loop level
\cite{MIZA, BEI1}. Its interpretation as a deformation of the spin chain 
Hamiltonian remains, however, an open question.

The relation between field theories and integrable models is 
not limited to very symmetric theories like ${\cal N}=4$
SYM. For example, orbifolds of ${\cal N}=4$ SYM in the planar limit
were discussed in \cite{chinese}. 
In the realm of non-supersymmetric theories, it was argued in
\cite{lipatov} that in certain QCD processes 
the summation of Feynman graphs leads to an integrable model
\cite{FAKO}. Following this line of reasoning, it was
shown (see e.g. \cite{BELI_QCD}) that various collections of operators
are closed under scale transformations and their anomalous dimensions
were determined. 

Given this amount of evidence it is interesting to ask which
4-dimensional field theories can exhibit an integrable structure in
the sense reviewed above. Some general observations immediately come
to mind. For example, planar  orbifolds of theories exhibiting an
integrable structure will continue to have one due to the ``inheritance 
principle'' \cite{BEJO}. Indeed, it was shown that in such theories
all planar diagrams are inherited from the parent theory, up to a
trivial rescaling of the gauge coupling. Also, deformations by
operators of dimension two do not affect the 1-loop scale
transformations and thus preserve whatever integrable structure the
undeformed theory might posses.

A complementary approach to the question formulated above asks which
integrable models can describe the 1-loop dilatation operator of some
field theory. If the field theory is in the planar limit, the choice 
is limited to models which have a lattice description with nearest 
neighbor interactions -- a spin chain.

In this note we will take a small step in this second direction. Starting
from a general integrable spin chain with only nearest neighbor
interactions we show that its Hamiltonian is the 1-loop dilatation
operator for some sector of a family of 4-dimensional field theories.
The general construction of the field theory associated to a given 
spin chain described in \S\ref{stupid} is straightforward and does 
not require the spin chain to have any
symmetry properties. Depending on the detailed properties of the spin
chain Hamiltonian, it is sometimes possible that the resulting
Lagrangian is part of a supersymmetric theory.
The construction implies a necessary condition for a given field
theory to be related to an integrable spin chain and suggests a way
to  attempt to engineer a spin chain once a field theory is
given. 

We then proceed to analyze in detail an example. With the eventual
goal of constructing a spin chain description of the 
q-deformation of the ${\cal N}=4$ SYM theory \cite{LEST}, \cite{BELE}, we
construct in \S\ref{bsc} an anisotropic spin chain with broken parity
invariance. Though we will be interested in a fixed number of states
per site, this number is kept arbitrary. We then construct its
associated field theory and find the constraints for it to be
supersymmetrizable. In \S\ref{LSD} we briefly review the 
q-deformation of the ${\cal N}=4$ SYM, show that it matches the
theory derived in \S\ref{bsc} and identify the sectors described by
the corresponding spin chain. 
In  \S\ref{BA} we use the Bethe Ansatz to diagonalize the spin chain
describing a 2-field sector of the theory. We also explicitly
construct some low dimension eigen-operators of the dilatation
operator. In \S\ref{end} we discuss possible extensions of this example.

\section{A trivial construction\label{stupid}}

Let us consider an one-dimensional lattice integrable model with
nearest neighbor interactions and periodic boundary 
conditions\footnote{The periodic boundary condition 
can easily be replaced with a twisted-periodic one. 
More general boundary conditions are also possible.}. 
Its Hamiltonian takes the usual form
\begin{equation}
H_J=\sum_{n=1}^J H_{n,n+1}~~~~~~~~~J+1\equiv 1
\label{hamilt}
\label{sc}
\end{equation}
where the index $n$ labels the chain site and
we assume that $H_{n,n+1}$ is independent of $n$. Each term in the sum
(\ref{hamilt}) acts in the obvious way:
\begin{equation}
H_{n,n+1}:V_n\otimes V_{n+1}\rightarrow V_n\otimes V_{n+1}
\end{equation}
and $V_i$ is the space of states at the $i$-th lattice site.

This data suffices to construct a 4-dimensional field theory whose
generator of scale transformations at 1-loop acts on a special 
class of scalar operators of bare dimension $J$ as $H_J$, 
up to the addition of the identity operator. The construction is
completely trivial. Consider the Lagrangian
\begin{equation}
{\cal L}=\sum_{i}Tr[\partial_\mu{\phi}_i
\partial^\mu{\bar \phi}^i]+\sum_{i,j,k,l}
Tr[\phi_i\phi_j{\bar \phi}^k{\bar\phi}^l ]
{\tilde H}_{kl}^{ji} + \sum_{i,j}Tr[\phi_j{\bar
\phi}^i\phi_l{\bar\phi}^k ]
{A}_{ik}^{jl}~~,
\label{Lc}
\end{equation}
where the fields $\phi$ are $n\times n$ matrices (with arbitrary $n$) and 
${\tilde H}$ and $A$ are, for the time being, arbitrary
coefficients. A simple computation implies that, at the planar
level, 1-loop scale transformations act on the holomorphic operators
with fewer than $n^2$ fields as
\begin{equation}
D=\frac{\lambda}{4\pi}
\sum_{i=1}^J\left[~{\tilde H}_{n_i,n_{i+1}}^{m_i,m_{i+1}}
+\alpha\delta_{n_i}^{m_i}\delta_{n_{i+1}}^{m_{i+1}}\right]~~.
\label{dil1}
\end{equation}
In deriving this equation we used dimensional regularization and
set all tadpole diagrams to zero.
The first term arises from the intrinsic renormalization of the
operator while the second term arises from the renormalization of the
constituent fields (the Lagrangian (\ref{Lc}) leads to $\alpha=0$, but
we keep $\alpha$ unspecified for further convenience).
 Thus, the spin chain (\ref{sc}) diagonalizes the
action of the dilatation operator on holomorphic operators in the
theory (\ref{Lc}) with
\begin{equation}
{\tilde H}=H_{n,n+1}~~,
\label{id1}
\end{equation}
while $A$ remains arbitrary. In fact, we can insert an arbitrary
multiplicative constant in the relation above. This coefficient will
appear only as a multiplicative factor between the eigenvalues of the
spin chain Hamiltonian and the anomalous dimensions of field theory
operators. 

We have therefore shown that, given any integrable spin chain, there
exists at least one field theory whose 1-loop dilatation operator acts
on certain operators as the Hamiltonian of the spin chain. 

Under certain conditions it is possible to construct field theories
such that the holomorphy constraint is absent. Indeed, if after
lowering the upper indices the spin chain Hamiltonian is cyclically
symmetric up to the addition of the identity operator
\begin{eqnarray}
&&H^{ji}_{kl}~~\longrightarrow~~H_{kl; ji}\\
H_{kl; ji}+\beta\delta_{kj}\delta_{li}&=&
H_{lj; ik}+\beta\delta_{li}\delta_{jk}=
H_{ji; kl}+\beta\delta_{jk}\delta_{il}=
H_{ik; lj}+\beta\delta_{il}\delta_{kj}\nonumber
\end{eqnarray}
then the computation leading to (\ref{dil1}) implies that the 1-loop
dilatation operator of the theory with Lagrangian
\begin{equation}
{\cal L}=\frac{1}{2}\sum_{i}Tr[\partial_\mu{\phi}^i
\partial^\mu{\phi}^i]+Tr[\phi^i\phi^j{\phi}^k{\phi}^l ]
({H}_{kl;ij}+\beta \delta_{kj}\delta_{li})
\label{Lr}
\end{equation}
is diagonalized by the eigenstates of the spin chain Hamiltonian. 

There are further theories, with similar properties, 
which differ from (\ref{Lc}) and (\ref{Lr})
by the addition of ``flavor-blind'' interactions. Such
interactions (e.g. gauge interactions) contribute only to the
coefficient $\alpha$ of the identity operator in (\ref{dil1}) and thus do
not modify the integrable properties of the 1-loop dilatation
operator.
From a field theory standpoint the different combinations of the terms
described
above lead to theories which are completely independent, though they
can be occasionally deformed into each other by adjusting the coupling
constants. It is however the case that the anomalous dimensions of
operators in appropriate sectors are not independent, but they are 
determined by the
eigenvalues of the transfer matrix of the same integrable spin chain.

\subsection{A necessary condition for 1-loop integrability}

As it is well known, a quantum integrable system is defined by its
R-matrix which is a solution of the quantum Yang-Baxter equation (QYBE)
\begin{equation}
R_{12}{}_{a_1a_2}^{b_1b_2}(\lambda-\mu)
R_{13}{}_{b_1a_3}^{c_1b_3}(\lambda)
R_{23}{}_{b_2b_3}^{c_2c_3}(\mu)
= 
R_{23}{}_{a_2a_3}^{b_2b_3}(\mu)
R_{13}{}_{a_1b_3}^{b_1c_3}(\lambda)
R_{12}{}_{b_1b_2}^{c_1c_2}(\lambda-\mu)~~.
\label{QYBE}
\end{equation}
The R-matrix determines the transfer matrix which can be used to
reconstruct the Hamiltonian of the system. More precisely, the 
logarithmic derivative of the transfer matrix evaluated at the value
of the spectral parameter for which the R-matrix becomes the 
permutation operator is a spin chain Hamiltonian with only nearest
neighbor interactions (\ref{hamilt}) with 
\begin{equation}
H_{n,n+1}\propto \frac{\partial
R}{\partial\lambda}\Big|_{\lambda=\lambda_0} ~~~~~~~~
R(\lambda_0)={\cal P}~~.
\end{equation}
Consequently, the discussion in the previous
section implies that the R-matrix of an integrable system is closely
related to the coefficients of the 4-point interaction terms in the
Lagrangian of the associated field theory. 

Finding solutions of the QYBE is in general a complicated task. The
simple observation made above yields a simpler (though still
algebraically rather involved) criterion for testing whether a given
4-point scalar interaction can be related to an integrable spin chain.

The idea is very simple: by taking one derivative of the QYBE with
respect to $\lambda$ and evaluating the result at specific values for
$\mu$ and $\lambda$ we obtain a constraint on $H$. The most obvious choice, $\lambda=\lambda_0=\mu$, leads to no useful constraints. A 
better choice turns out to be
\begin{equation}
\lambda\rightarrow\infty
~~~~~~~~
\mu\rightarrow\infty
~~~~~~~~
\lambda-\mu=\lambda_0~~.
\end{equation}
This limit is easy to analyze because for infinite spectral
parameter the R-matrix becomes
\begin{equation}
R(\lambda)=\id\otimes\id + \frac{r}{\lambda}+{\cal O}(1/\lambda^2)
\end{equation}
where $r$ is a solution of the classical Yang-Baxter equation:
\begin{equation}
[r_{12},\,r_{13}]+[r_{12},\,r_{23}]+[r_{13},\,r_{23}]=0~~.
\end{equation}
Similarly to $R_{ij}$, 
$r_{ij}$ acts on $V_i\otimes V_j$ and the commutators are taken
over repeated indices.

Interpreting $H_{12}\equiv H_{i_1i_2}^{j_1j_2}$ as an operator acting
on $V_1\otimes V_2$, simple algebra leads to the conclusion that, for 
$H_{12}$ to be related to an integrable spin chain, it is
necessary that 
\begin{equation}
[{\cal P}H_{12},\,r_{23}]+ [{\cal P}H_{12},\, r_{13}]=0
\label{constraint}
\end{equation}
where $r$ is a solution of the classical Yang-Baxter equation and 
$({\cal P}H){}_{ab}^{cd}=H{}_{ba}^{cd}$.
It is worth pointing out that the 
equation (\ref{constraint}) is insensitive to
the freedom of adding the identity operator in the relation between 
the spin chain and the 4-point coupling in the field theory Lagrangian. 

Thus, the conclusion is that for a given field theory with only
4-point interactions to have a chance to be related to an integrable
spin chain, the equation (\ref{constraint}) must be satisfied with
$H_{12}$ being the coefficient of the 4-point coupling and $r$ being a
solution of the classical Yang-Baxter equation. As an example let us
briefly ilustrate this constraint with the 
holomorphic sector of ${\cal N}=4$ SYM. In this case, $H$ is a linear
combination of the identity operator and permutation
operator.
The
equation (\ref{constraint}) is satisfied if $r$ is also the
permutation operator, which also satisfies the classical Yang-Baxter
equation.

\subsection{Symmetries}

The Lagrangians (\ref{Lc}) and (\ref{Lr}) have the same symmetries as
the Hamiltonian of the  
integrable spin chain they are based on. Some integrable spin chains 
have Hamiltonians which are invariant under the action of some group 
$G$. In such situations the fields $\phi^i$ transform in some 
representation of the symmetry group. This is, however, the exception
rather than the rule, since probably most integrable models have no
nonabelian continuous bosonic symmetries. In \S\ref{bsc} we will
construct such a model and analyze it in detail.

An interesting question is whether the Lagrangian (\ref{Lc}) can be
embedded in a supersymmetric theory. It is easy to see that fermions
can be added to (\ref{Lc}) without spoiling the integrability of
(\ref{dil1}). Indeed, Lorentz invariance guarantees that at 1-loop
level the only contribution of fermions to the scale transformation of
bosonic scalar operators is through the wave function renormalization
and thus affects only the precise value of the coefficient $\alpha$ in
(\ref{dil1}).  

Since the scalar fields in chiral multiplets are complex, it is
natural to start with a field theory Lagrangian of the type (\ref{Lc}).
Requiring that it is supersymmetrizable imposes certain factorization 
constraints on the spin chain Hamiltonian (\ref{sc}), since the
interaction term in (\ref{Lc}) must be positive semidefinite. Thus, we
find that $H_{n, n+1}$ must factorize as
\begin{equation}
H_{n,n+1}{}_{kl}^{ji}=\sum_k\sum_{M_k} C_{M_k\,kl}{\bar C}^{M_k\,ji}
\end{equation}
where the bar denotes complex conjugation and the ranges of the indices
$k$ and $M_k$ are fixed by the solution of the factorization problem.

Even if $H$ has this property, the generic situation is that
the sum of the ranges of $M_k$ exceeds the range of $i$ and $j$.
Then, the Lagrangian (\ref{Lc}) can be only a subsector of a larger
theory, since it is necessary to add at least $\sum_k|M_k|-n$ extra
fields. If this difference vanishes or equals unity the theory is
special in the sense that, depending on $C$, it may not require extra
fields for it to be supersymmetric (vanishing difference) or just one
extra field which may be a gauge field (unit difference). 

A full classification is cumbersome (and perhaps not very illuminating) 
without making further assumptions on $H$ and $C$. We will therefore
leave the general discussion and proceed to analyze in detail a
specific example. In the next section we construct an integrable spin
chain which is a deformation of one with $U(K)$ invariance. 
In general, the resulting chain has no continuous nonabelian
symmetries. For special values of the 
coupling constants the associated field theory can be embedded in a
supersymmetric one. It will turn out that, if each site supports two
states ($K=2$), this chain describes the holomorphic operators in a  
2-field sector of a q-deformation of ${\cal N}=4$ SYM
with arbitrary deformation parameter while for three states per site 
it describes all
holomorphic  operators  of the q-deformation of ${\cal
N}=4$ SYM with a pure phase parameter.

\section{An anisotropic spin chain with broken parity \label{bsc}}

As an example of the previous discussion, in the remainder of this
note we will describe in the language of
integrable spin chains (part of) the holomorphic sector of
certain deformations of ${\cal N}=4$ SYM
theory which break its R-symmetry to a $U(1)^3$ subgroup. 

Since the R-symmetry of ${\cal N}=4$ SYM restricted to the
holomorphic sector is $SU(3)$ we should in principle look for
R-matrices which act on the 3-dimensional representation of $SU(3)$
but do not posses this symmetry. We will however keep the discussion
general and construct an R-matrix acting on the fundamental
representation of $U(K)$. The advantage of doing this is that the
result can be used to describe smaller sectors of the field theory.
We will analyze in considerable detail a 2-field sector of the theory
and find the spectrum of scaling dimensions of all chiral operators in
this sector.

\subsection{The R-matrix and the spin chain Hamiltonian}

Parameter-dependent R-matrices are a common feature in the theory of
quantum groups. Besides the spectral parameter $\lambda$, R also
depends on the quantum deformation parameter $q$. There also exist
quantum groups with more than one deformation parameter. Their
associated R-matrices will depend on these parameters and are clearly
not invariant under the undeformed symmetry transformations. Even more
general parameter-dependent R-matrices can be constructed.

Non-symmetric solutions of the quantum Yang-Baxter equation 
(\ref{QYBE})
are usually constructed on a case by case basis. Defining 
$(e^{ij})_{kl}=\delta^i_k\delta^j_l$ with $i, j, k, l=1,\dots, K$, 
it is easy though rather tedious to check that a solution of
(\ref{QYBE}) is 
\begin{eqnarray}
R(\lambda)&=&\left(q^{1+a \lambda}-q^{-1-b
\lambda}\right)\sum_ie^{ii}\otimes 
e^{ii}+\left(q^{a \lambda}-q^{-b \lambda}\right)\sum_{i\ne
j}e^{\alpha_{ij}}e^{ii}\otimes e^{jj} \nonumber\\
&+& (q-q^{-1})\left(q^{a \lambda}\sum_{i< j}e^{ij}\otimes
e^{ji}+q^{-b \lambda}\sum_{i> j}e^{ij}\otimes
e^{ji}\right)
\label{Rmat}
\end{eqnarray}
where $\lambda$ is the spectral parameter while $a$, $b$, $q$ and
$\alpha_{ij}=-\alpha_{ji}$ are free parameters. Similar R-matrices
appeared in \cite{BMRA, NABA}. 
This R-matrix has all the properties described above and we
will take it as the starting point of our construction.

Choosing the Lax operator to be the R-matrix, the monodromy matrix
$T$, which is a solution of the equation 
\begin{equation}
R_{12}(\lambda-\mu)(T(\lambda)\otimes T(\mu)) = 
(T(\mu) \otimes T(\lambda))R_{12}(\lambda-\mu)
\label{QYBET}
\end{equation}
and the transfer matrix are constructed out of (\ref{Rmat}) in the
usual way:
\begin{eqnarray}
T(\lambda)_{~\,a_1\, ;i_1\dots i_n}^{a_{n+1};j_1\dots j_n} 
&=&\sum_{a_2,\dots a_n} 
R_{a_1i_1}^{a_2 j_1}R_{a_2i_2}^{a_3 j_2}\dots   
R_{a_{n-1}i_{n-1}}^{a_n j_{n-1}}R_{a_ni_n}^{a_{n+1} j_n}\label{T}\\
\tau(\lambda)_{i_1\dots i_n}^{j_1\dots j_n}&=&
\sum_{a}T(\lambda)_{a ;i_1\dots i_n}^{a;j_1\dots j_n}~~.
\end{eqnarray}

In the construction of the Hamiltonian, an important value of the
spectral parameter is the one for which the R-matrix becomes the
permutation operator and the monodromy matrix becomes the shift
operator. In our case, this value is $\lambda=0$
\begin{equation}
R(0)=(q-q^{-1})\sum_{i,\,j}e^{ij}\otimes e^{ji} = (q-q^{-1}){\cal
P}~~~~~~~ {\cal P}_{i,j}^{k,l}=\delta_i^l\delta_j^k~~.
\label{special}
\end{equation}
It is then easy to see that the transfer matrix evaluated at vanishing
spectral parameter is the generator of shifts along the chain. 

The Hamiltonian of the spin chain is the logarithmic derivative of 
the transfer matrix evaluated (for the present case) at $\lambda=0$. 
The equation (\ref{special}) implies that it
is a sum of nearest neighbor interaction terms, each of which is, up
to normalization,  the derivative of the R-matrix evaluated at
$\lambda=0$: 
\begin{eqnarray}
H_{n,n+1}
&=&-b\ln
q\left[ \frac{1-\Delta^2 q^2}{1-q^2}
\sum_{i}e_{n}^{ii}\otimes e_{n+1}^{ii} 
+ \frac{q(1-\Delta^2)}{1-q^2}
\sum_{i\ne j}e^{\alpha_{ij}}
e_{n}^{ij}\otimes e_{n+1}^{ji} \right.\nonumber\\
 && ~~~~~~~~+\left.
\Delta^2 \sum_{i< j}e_{n}^{ii}\otimes e_{n+1}^{jj} 
+ \sum_{i> j}e_{n}^{ii}\otimes e_{n+1}^{jj} \right]
\label{sch}
\end{eqnarray}
where we defined $\Delta^2=-a/b$.

We are now ready to construct the field theory associated to this spin
chain. In the construction above the range of the indices of the
R-matrix was not fixed. In the following sections however we will be
interested in specific cases in which $i$ and $j$ take two or three
values\footnote{Alternatively, $e^{ij}$ generate $U(2)$ and $U(3)$,
respectively.}.

\subsection{The associated field theory}

To illustrate the construction in \S\ref{stupid}
we will write down the associated field theory first for $e^{ij}$
generating $U(2)$ and then for arbitrary range for its indices.

In the first case $e^{ij}$ can be expressed in terms of the usual
Pauli matrices
\begin{equation}
e^{12}=\sigma^+~~~~~~~~e^{21}=\sigma^-~~~~~~~~
e^{11}=\frac{1}{2}\left(\id+\sigma^3\right)~~~~~~~~
e^{22}=\frac{1}{2}\left(\id-\sigma^3\right)~~.
\label{pauli}
\end{equation}
The resulting spin chain is a parity-violating\footnote{For $a\ne -b$
the interactions to the left are different from those to the right.} 
extension of the  XXZ Heisenberg chain. The latter is recovered in the
limit $q\rightarrow 1,~\alpha\rightarrow 0$ and $a\rightarrow -b$
taken in this order at the level of the R-matrix.  It is easy to see
from (\ref{sch}) that this extension does not include the XYZ spin chain.
As in the case of the XYZ chain, the natural $U(2)$ symmetry of the
system is broken by 
the presence of the various relative coefficients in (\ref{sch}).

From the construction in the previous section, a field theory
associated to this spin chain has the following Lagrangian:
\begin{eqnarray}
\label{ex_bose}
&&{\cal L}=Tr\left[\partial\phi_i\partial{\bar\phi}^i-
F\,b\ln q\left[ \frac{q(1-\Delta^2)}{1-q^2}\left(e^\alpha
\phi_2\phi_1{\bar\phi}^2{\bar\phi}^1
+
e^{-\alpha}\phi_1\phi_2{\bar\phi}^1{\bar\phi}^2\right)\right.\right.
\nonumber\\
&+&\left.\left.
\Delta^2 \phi_2\phi_1{\bar\phi}^1{\bar\phi}^2
+\phi_1\phi_2{\bar\phi}^2{\bar\phi}^1+
\frac{1-\Delta^2 q^2}{1- q^2}\left(\phi_1\phi_1{\bar\phi}^1{\bar\phi}^1
+\phi_2\phi_2{\bar\phi}^2{\bar\phi}^2\right)\right]\right]
\end{eqnarray}
where the coefficients identify the origin of each of the terms
above in equation (\ref{sch})
and the coefficient $F$ reflects the freedom to rescale the spin
chain Hamiltonian. 

It is also not possible to extend (\ref{ex_bose}) to a supersymmetric
Lagrangian without adding at least three more bosonic fields. As it is
well known, a theory admits a supersymmetric extension if its potential
is a sum of squares. There are many ways one can attempt to arrange
the terms in (\ref{ex_bose}) in this fashion. Furthermore, as pointed
out in \S\ref{stupid}, there also exists the freedom of adding terms
which do not contribute to planar Feynman diagrams. Keeping this in
mind, a useful way of writing (\ref{ex_bose}) is
\begin{eqnarray}
{\cal L}=\partial\phi_i\partial{\bar\phi}^i\! &-&\!
F\,\ln q\left[\, k|\phi_1\phi_2 - W\,\phi_2\phi_1|^2+ C\,
(\phi_1{\bar\phi}^1+\phi_2{\bar\phi}^2)(
{\bar\phi}^1{\phi}_1+{\bar\phi}^2{\phi}_2)\right.\nonumber\\
&+&\!\left.
A
\phi_2\phi_1{\bar\phi}^2{\bar\phi}^1
+
B
\phi_1\phi_2{\bar\phi}^1{\bar\phi}^2
\right]
\label{qftlag}
\end{eqnarray}
where
\begin{eqnarray}
\nonumber
A=kW+b\frac{q(1-\Delta^2)}{1-q^2}e^\alpha
~~~~
B\!\!&=&\!\!k{\overline W}+b\frac{q(1-\Delta^2)}{1-q^2}e^{-\alpha}
~~~~
C=b\,\frac{1-\Delta^2 q^2}{1- q^2}
\\
k=-b\frac{q^2\,(1-\Delta^2)}{1-q^2}&&~~~~~~~~|W|^2=\frac{1}{q^2}
\label{constr1}
\end{eqnarray}
By adding terms contributing only to nonplanar (and perhaps
self-energy) diagrams, the last term on the first line can be written
as a perfect square of an imaginary object. 
Nevertheless, the Lagrangian (\ref{qftlag}) cannot be supersymmetrized
as long as $A\ne 0$ and $B\ne 0$. It may already be clear and we will 
see it in detail in \S\ref{LSD} that if $A$ and $B$ vanish, 
this theory is (part of) the q-deformation of the ${\cal
N}=4$ SYM.

In the general case, the interaction Lagrangian of the field theory
associated to the spin chain (\ref{sch}) is just
\begin{eqnarray}
\label{gen_qftlag}
\frac{L_{\it int}}{F \ln q}
&=&C\left(\sum_{i=1}^n \phi_i {\bar\phi}^i\right)
\left(\sum_{i=1}^n {\bar \phi}^i {\phi}_i\right)
+\sum_{i<j=1}^n k_{ij}|\phi_i\phi_j-W_{ij}\phi_j\phi_i|^2 \\
&+&
\sum_{i<j=1}^n A_{ij}\phi_j\phi_i{\bar \phi}_j{\bar\phi}_i
+
B_{ij}\phi_i\phi_j{\bar \phi}_i{\bar\phi}_i~~.
 \nonumber
\end{eqnarray}
It turns out that $k_{ij}=k$ and $|W_{ij}|=|W|$ for all indices $i,j=1\dots n$
while $A_{ij}$, $B_{ij}$ and the phase of $W_{ij}$ can be obtained from
(\ref{constr1}) by replacing $\alpha$ with $\alpha_{ij}$.
As before, if the terms on the second line vanish, this Lagrangian
can be supersymmetrized by adding an appropriate number of fields as
well as the appropriate ``nonplanar'' terms.

It is important to note that the coefficient $C$ in the equations
(\ref{qftlag}) and (\ref{gen_qftlag}) is not fixed.
Indeed, we are free to add the identity operator to
(\ref{sch}) without changing the eigenvectors and shifting the
eigenvalues by its coefficient. This freedom will be important in the
next section. 

\section{The q-deformation of ${\cal N}=4$ SYM \label{LSD}}

In this section we will show that the supersymmetric limit of the
Lagrangians (\ref{qftlag}) and (\ref{gen_qftlag}) describe,
respectively, a 2-field sector of a  q-deformation with
generic deformation parameter of the ${\cal N}=4$ SYM and the full
holomorphic sector if the deformation parameter is a pure phase.

There is no {\it a priori} reason to expect that deformations of 
${\cal N}=4$ SYM should have integrable structures. 
Various sectors of ${\cal N}=4$ SYM are described by
certain integrable spin chains and some of them have deformations which
preserve the integrability. Usually, such deformations are, however, 
rather difficult to find. The spin chain constructed in \S\ref{bsc} is
an integrable deformation of the $U(K)$ symmetric chain in the
fundamental representation.

Let us begin by reviewing the q-deformation of the 
${\cal N}=4$ SYM. The superpotential is given by
\begin{equation}
W=Tr[\Phi_1(\Phi_2\Phi_3-w\Phi_3\Phi_2)]~~. 
\end{equation}
This deformation preserves ${\cal N}=1$ supersymmetry and breaks the
$SU(4)$ R-symmetry to $U(1)^3$. 
Integrating out the
auxiliary fields and restricting to the scalar sector the interaction
terms are
\begin{eqnarray}
V&=&Tr[~|\phi_1\phi_2 -w\phi_2\phi_1|^2+|\phi_2\phi_3 -w\phi_3\phi_2|^2
+ |\phi_3\phi_1 -w\phi_1\phi_3|^2 ]\nonumber\\
&+&Tr[~([\phi_1,\,{\bar \phi}^1]+[\phi_2,\,{\bar
\phi}^2]+[\phi_3,\,{\bar \phi}^3])^2] 
\label{vv}
\end{eqnarray}
where the first three terms arise from integrating out the auxiliary
fields in the chiral multiplets while the last term arises from 
integrating out the auxiliary fields in the vector multiplet.

A brief look at the appropriate Feynman diagrams shows that, as in 
the case of ${\cal N}=4$ SYM ($w=1$), operators built out of only 
two scalar fields form a sector closed under scale transformations. 
Furthermore, due to the large $N$ limit, the terms contributing to
the 1-loop scale transformations of these operators are obtained 
by setting one of the fields (say $\phi_3$) to zero as well as keeping
only the terms in which the conjugate fields appear next to each other
\begin{eqnarray}
V&=&Tr[~|\phi_1\phi_2 -w\phi_2\phi_1|^2]-2Tr[~(\phi_1{\bar
\phi}^1+\phi_2{\bar\phi}^2)({\bar \phi}^1 {\phi}_1+
{\bar\phi}^2{ \phi}_2)] ~~.
\label{2fs}
\end{eqnarray}
Comparing this expression with the supersymmetric limit of
(\ref{qftlag}) we find that they are identical provided that we set
\begin{eqnarray}
&&A=0~~~~~~~~~~~B=0~~~~\Rightarrow~~~~W=\frac{e^{i\beta}}{q}~~~~\alpha=i\beta
~~~~\beta\in\Re\\
&&w=W~~~~~~~~F\,k\,\ln q=1~~~~~~~~\frac{C}{k}=
-\frac{1-\Delta^2q^2}{q^2(1-\Delta^2)} =-2~~.
\label{constraints_2}
\end{eqnarray}
The first four identifications are rather straightforward. The fifth
one fixes $\Delta$ in terms of the absolute value of $w$. It is worth
pointing out that all the free parameters appearing in the Hamiltonian
(\ref{sch}) were necessary to recover the details of (\ref{2fs}).

As pointed out in the previous section, the relative coefficient
between the two terms in (\ref{2fs}) has no consequence in this match,
since it is possible to adjust it by adding the identity operator to
the spin chain Hamiltonian. The holomorphic
operators in the 2-field sector have the following structure
\begin{equation}
{\cal O}=Tr[\prod_{i=1}^n\phi_1^{m_i}\phi_2^{p_i}]
\label{2_field_op}
\end{equation}
with arbitrary $n$, $m_i$ and $p_i$. We will construct the eigenstates
of the generator of scale transformations in \S\ref{esev}. In the
limit $w\rightarrow 1$ these operators reduce to those discussed in 
\cite{BMSZ}.

The three field sector can be matched with the Lagrangian
(\ref{gen_qftlag}) for $n=3$. In this case however there is a further
constraint which arises because of the different order of fields in
(\ref{gen_qftlag}). Indeed, after adding terms which do not contribute
in the planar limit, adjusting the coefficients $F$ and $C$ 
in (\ref{gen_qftlag}) and taking the supersymmetric limit, 
the Lagrangian associated to the 3-state spin chain becomes
\begin{eqnarray}
V
&=&Tr([\phi_1,\, {\bar\phi}^1]+
[\phi_2,\, {\bar\phi}^2]+[\phi_3,\, {\bar\phi}^3])^2  \\
&+&Tr|\phi_1\phi_2-W_{12}\phi_2\phi_1|^2 +
|\phi_2\phi_3- W_{23}\phi_3\phi_2|^2 
+|\phi_1\phi_3-W_{13}\phi_3\phi_1|^2~~. \nonumber
\end{eqnarray}
where $W_{ij}=q^{-1}e^{i\beta_{ij}}$ with $\alpha_{ij}=i\beta_{ij}$,
similar to the two-field case. Thus, it matches the terms in (\ref{vv})
contributing to the 1-loop scale transformations of holomorphic operators 
\begin{equation}
{\cal
O}=Tr[\prod_{i=1}^n\phi_1^{m_i}\phi_2^{p_i}\phi_3^{q_i}]~~~~~(\forall) 
~~n, ~m_i,~p_i~{\rm and }~q_i
\end{equation}
only if $w=W_{12}=W_{23}={\overline W}_{13}$ is a phase. This
constraint amounts to taking the limit $q\rightarrow 1$ at the level
of the Hamiltonian (\ref{sch}). This is a rather singular limit, which
should be taken after the supersymmmetry conditions $A_{ij}=B_{ij}=0$
are imposed. It may be possible to relax the constraint that $|w|=1$ by
considering an even more general R-matrix.  
Finding the eigenvectors of the generator of scale transformations
requires the diagonalization of the 3-state spin chain Hamiltonian,
which can be achieved through a nested Bethe Ansatz.

\section{ The Bethe Ansatz in the 2-field sector \label{BA}}

We will now diagonalize the Hamiltonian (\ref{sch}) under the
assumption that $(e^{ij})$ generate an $SU(2)$ algebra. Since a
systematic construction of the eigenstates relies on the details of
the diagonalization procedure, we will describe it in some detail.
Following \cite{KOBI}, we will use the Algebraic Bethe
Ansatz. This method also applies to spins with more than two
states per site, as it was discussed in \cite{NABA} for slightly
different though very similar  Hamiltonians. 

To fix the notation, if $(e^{ij})$ generate an $SU(2)$ algebra in the
fundamental representation, the monodromy matrix acts on a
two-dimensional auxiliary space; its entries are operators 
which act on the quantum states
\begin{equation}
T(\lambda)=\pmatrix{A(\lambda) & B(\lambda) \cr C(\lambda) &
D(\lambda)}~~.
\label{mono}
\end{equation}
If the spin chain has $J$ sites, the operators 
$A,\,B,\,C$ and $D$ act on the tensor product of $J$ copies of the
fundamental representation of $SU(2)$, in spite of the fact that they
are not $SU(2)$ invariants. From (\ref{mono}) it follows that the 
transfer matrix is
\begin{equation}
\tau(\lambda)=Tr_{\it aux}T(\lambda)=A(\lambda)+D(\lambda)~~.
\end{equation}

We will consider the following general R-matrix 
\begin{equation}
R(\lambda-\mu)=\pmatrix{f_u(\lambda,\mu) & 0          & 0          & 0\cr
		    0          & g_u(\lambda,\mu) & h(\lambda,\mu) & 0\cr
		    0          & k(\lambda,\mu) & g_l(\lambda,\mu) & 0\cr
		    0          & 0          & 0          &
f_l(\lambda,\mu) }~~,
\label{genR}
\end{equation}
with $f_u(\lambda,\mu) =f_u(\lambda-\mu)$, etc.
In this section we will not need the specific forms of $f, ~g,~h$ and 
$k$. They are fixed by the QYBE, up to an overall multiplicative
function of $\lambda$. Clearly, (\ref{Rmat}) is of this type.

The equation (\ref{QYBET}) implies certain commutation relations
between the entries of the monodromy matrix evaluated at different
values for the spectral parameter. For the present discussion, the
most important ones are
\begin{eqnarray}
&&[A(\lambda),\,A(\mu)]=0~~~~~~~~
[B(\lambda),\,B(\mu)]=0\label{com0}\\
&&[C(\lambda),\,C(\mu)]=0~~~~~~~~
[D(\lambda),\,D(\mu)]=0\\
&&
A(\mu)B(\lambda)=
\frac{f_u(\lambda,\mu)}{g_l(\lambda,\mu)}B(\lambda)A(\mu)
-\frac{h(\lambda,\mu)}{g_l(\lambda,\mu)}B(\mu)A(\lambda)\label{com2}\\
&&
D(\mu)B(\lambda)=
\frac{f_l(\mu,\lambda)}{g_l(\mu,\lambda)}B(\lambda)D(\mu)
-\frac{k(\mu,\lambda)}{g_l(\mu,\lambda)}B(\mu)D(\lambda)
\label{com3}
\end{eqnarray}
which suggest that $B$ and $C$ can be interpreted as creation and
annihilation operators. Thus, using a pseudo-vacuum 
state $|0\rangle$ satisfying
\begin{equation}
A(\lambda)|0\rangle = a(\lambda) |0\rangle~~~~~~~~
D(\lambda)|0\rangle = d(\lambda) |0\rangle~~~~~~~~
C(\lambda)|0\rangle = 0
\label{vacuum}
\end{equation}
the ansatz for the level $N$ eigenstates of the transfer matrix is:
\begin{equation}
|\Psi_N(\{\lambda_i\})\rangle=\prod_{i=1}^N B(\lambda_i)|0\rangle~~.
\label{ansatz}
\end{equation}

Requiring that (\ref{ansatz}) is indeed an eigenstate of the transfer
matrix leads to the Bethe equations, which constrain the arguments
$\lambda_i$ of the creation operators appearing in
$|\Psi_N(\{\lambda_i\})\rangle$. Using (\ref{com0})-(\ref{com3}) 
it is fairly easy to find that
\begin{eqnarray}
&&A(\mu)|\Psi_(\{\lambda_i\})\rangle=
\Lambda|\Psi_N(\{\lambda_i\})\rangle+
\sum_{n=1}^N \Lambda_n B(\mu) \prod_{\stackrel{i=1}{i\ne n}}^N
B(\lambda_i)|0\rangle
\label{up}\\
&&D(\mu)|\Psi_N(\{\lambda_i\})\rangle=
{\tilde \Lambda}|\Psi_N(\{\lambda_i\})\rangle+
\sum_{n=1}^N{\tilde \Lambda}_n B(\mu) \prod_{\stackrel{i=1}{i\ne n}}^N
B(\lambda_i)|0\rangle
\label{down}
\end{eqnarray}
with the coefficient functions $\Lambda$ given by
\begin{eqnarray}
\Lambda=
a(\mu)\prod_{i=1}^N\frac{f_u(\lambda_i, \mu)}{g_l(\lambda_i, \mu)}
~
&&{\tilde\Lambda}=d(\mu)\prod_{i=1}^N\frac{f_l(\mu,\lambda_i)}
{g_l(\mu,\lambda_i)} \\
\Lambda_n=-a(\lambda_n)\frac{h(\lambda_n,\mu)}{g_l(\lambda_n,\mu)}
\prod_{\stackrel{i=1}{i\ne
n}}^N\frac{f_u(\lambda_i, \lambda_n)}{g_l(\lambda_i, \lambda_n)}
~
&&{\tilde\Lambda}_n=-d(\lambda_n)
\frac{k(\mu,\lambda_n)}{g_l(\mu,\lambda_n)}
\prod_{\stackrel{i=1}{i\ne
n}}^N\frac{f_l(\lambda_n,\lambda_i)}{g_l(\lambda_n,\lambda_i)}
\nonumber
\end{eqnarray}
In the equations above $\Lambda$ and ${\tilde \Lambda}$ arise from
using the first term in (\ref{com2}) and (\ref{com3}), respectively.
The other $(2^J-1)$ terms, coming from commuting $A$ past the $J$
factors of $B$, combine into $\Lambda_n$ and ${\tilde \Lambda}_n$. To
see that the expressions quoted above are correct we first note that 
(\ref{ansatz}) is completely symmetric in $\lambda_i$ which implies
that it is
enough to compute $\Lambda_1$ and ${\tilde \Lambda}_1$. 
They arise from using the second term in (\ref{com2}) and
(\ref{com3}) for pushing $A$ and $D$ past $B(\lambda_1)$ and the first
term in those equations for the remaining  commutators. 

The equations (\ref{up}) and (\ref{down}) imply that the vectors
(\ref{ansatz}) are eigenvectors of the transfer matrix  if the terms
depending on $B(\mu)$ cancel out. This condition 
leads to $N$ equations which
determine the arguments of the creation operators in terms of $\mu$:
\begin{equation}
\Lambda_n+{\tilde
\Lambda}_n=0~~~~~\leftrightarrow~~~~~~
\frac{a(\lambda_n)}
{d(\lambda_n)} 
= \prod_{\stackrel{i=1}{i\ne
n}}^N\frac{f_l(\lambda_n,\lambda_i)}{f_u(\lambda_i,\lambda_n)}
\frac{g_l(\lambda_i,\lambda_n)}{g_l(\lambda_n,\lambda_i)}
\label{BE}
\end{equation}
where we used the identity
\begin{equation}
\frac{g_l(\mu,\lambda_n)h(\lambda_n,\mu)}
{g_l(\lambda_n\mu)k(\mu,\lambda_n)} =-1
\end{equation}
which is a consequence of the QYBE.

These are the Bethe equations. Interestingly enough, these equations
do not depend directly on the off-diagonal entries of the
R-matrix. 
The fact that the R-matrix is not symmetric is reflected by 
the lack of a simple relation between
$g_l(\lambda_i,\lambda_j)$ and $g_l(\lambda_j,\lambda_i)$ (which are
``usually'' the negative of each other).

There is an additional constraint on the arguments $\{\lambda_i\}$ of
the creation operators $B$, which arises from the fact that we want to
associate the eigenvectors of $t$ with single-trace operators. The
cyclicity of the trace translates into the constraint that 
(\ref{ansatz}) is invariant under shift operator $t(0)$:
\begin{equation}
 a(0)\prod_{i=1}^N\frac{f_u(\lambda_i)}{g_l(\lambda_i)}
+d(0)\prod_{i=1}^N\frac{f_l(-\lambda_i)}{g_l(-\lambda_i)}=1
\end{equation}

Out of the eigenvalues of the transfer matrix it is now trivial to
construct the eigenvalues of the Hamiltonian, by taking the
appropriate logarithmic derivative with respect to $\mu$ and
evaluating the result at $\mu=0$, as we did in (\ref{sch}) to
construct the Hamiltonian:
\begin{equation}
E(\{\lambda_i\})=\frac{\partial\ln(\Lambda+{\tilde\Lambda})}{\partial\mu}
\Big|_{\mu=0}=\epsilon_0+\sum_{i=1}^N\,\epsilon_i~~.
\end{equation}
In the equation above $\epsilon_0$ arises from derivatives acting 
on $a(\mu)$ and $d(\mu)$, while $\epsilon_i$
contains the terms in which the derivative acts on the factor
depending on $\lambda_i$ in $\Lambda$ and ${\tilde \Lambda}$. In some
sense, $\epsilon_0$ can be thought as some ``vacuum energy'', since it
is proportional to the number of sites in the chain; its precise value
depends on the normalization of the R-matrix. In the same spirit,
$\epsilon_i$ can be thought as the contribution of a single creation
operator to the energy.

It is now a simple exercise to apply these results to 
the case of the spin chain (\ref{sch}) and find the 1-loop
anomalous dimensions of the holomorphic operators (\ref{2_field_op})
in the theory (\ref{qftlag}).
Since the supersymmetric case can be obtained from the
non-supersymmetric one by a specific choice of parameters, we will
keep the discussion general.

The diagonalization of the 3-state spin chain Hamiltonian can be
achieved through the nested Bethe Ansatz. The analysis is fairly 
similar to the
2-state case, though substantially more involved. For the spin chain
corresponding to the Izergin-Korepin R-matrix, this was described in
detail in \cite{REVI}, while spin chains similar to those in
\S\ref{bsc} were analyzed in \cite{NABA}.

\section{The eigenstates and eigenvalues \label{esev}}

We will now specify the solution discussed in \S\ref{BA} to the spin
chain constructed in \S\ref{bsc}. 
In this case the entries of the R-matrix are ($f(\lambda)\equiv
f(\lambda, 0)$, etc.):
\begin{eqnarray}
f_u(\lambda)=f_l(\lambda)&=&q^{1+a\lambda}-q^{-1-b\lambda}\\
g_u(\lambda)=e^\alpha\left[q^{a\lambda} - q^{-b\lambda}\right]
~~~&&~~~
g_l(\lambda)=e^{-\alpha}\left[q^{a\lambda} - q^{-b\lambda}\right]\\
h(\lambda)=(q^2-1)q^{a\lambda-1}
~~~&&~~
k(\lambda)=(q^2-1)q^{-b\lambda-1}
\label{rmatentries}
\end{eqnarray}

The structure of the R-matrix implies that on each site the $C$
operator acts as $\sigma^+$. Thus, the vacuum state $|0\rangle$
corresponds to the operator
\begin{equation}
O^J_0=Tr[\phi_1^J]\equiv | \underbrace{\uparrow\uparrow\dots
\uparrow}_{J~{\rm times}}  \,\rangle
\label{vacuum_ex}
\end{equation}
where $J$ is the number of sites in the chain. In the ket vector
notation the up and down arrows correspond  to $\phi^1$ and $\phi^2$,
respectively. Thus, the arrows are cyclically-symmetric. This is
important for correctly computing their normalization.

From this choice of the vacuum state 
it is easy to see that $a(\mu)$ and $d(\mu)$ in
(\ref{vacuum}) are given by
\begin{equation}
a(\mu)=f_u(\mu)^J~~~~~~~d(\mu)=g_l(\mu)^J
\label{vacen}
\end{equation} 
Then, the vacuum and creation operator contribution to the 
eigenvalues of the transfer matrix are:
\begin{equation}
\epsilon_0 = J \, \frac{a q+b/q}{q-1/q}\,\ln q
~~~~~~~~
\epsilon_i=
\frac{(a+b)(q-1/q)\ln
q}{(1-q^{-(a+b)\lambda_i})(q^{1+(a+b)\lambda_i}-1/q)}~~.
\label{energ_parts}
\end{equation}
Using (\ref{rmatentries}) and (\ref{vacen}) in (\ref{BE}) 
and taking the limit
$q\rightarrow 1$ and $\alpha\rightarrow 0$ as well as shifting
$\lambda$ by a constant recovers the classic equations for the
XXZ spin chain\footnote{From the equation (\ref{energ_parts}) it is
clear that the order of limits needed to recover the XXZ spin chain is
the one stated below equation (\ref{pauli}): 
$q\rightarrow 1,~\alpha\rightarrow 0$ and $a\rightarrow -b$}. 

The anomalous dimensions of the linear combinations of operators
(\ref{2_field_op}) corresponding to the eigenstates (\ref{ansatz})
are obtained from (\ref{energ_parts}) by using the parameters 
$F,\,a,\,b$ and $q$ determined by (\ref{constraints_2}) as well as adding
the contribution of the wave function renormalization. It is not hard
to see that this leads to a vanishing anomalous dimension for the
ground state (\ref{vacuum_ex}), as it should. It is interesting to note
that this result holds even if supersymmetry is broken by having
$A\ne0$ or/and $B\ne 0$. This can be easily checked from field theory
considerations by noticing that
the susy-breaking terms in (\ref{qftlag}) do not contribute to the
1-loop scale transformation of the operator dual to the ground state
(\ref{vacuum}).

The eigenvectors are constructed by acting with the creation operators
$B(\lambda)$ on $O_0^J$. The expression of $B$ for arbitrary $J$ is
rather complicated and perhaps not very enlightening. 
Roughly speaking, the action of a single creation operator creates a
``spin wave''. Consequently, the eigen-operators have a structure
similar to that of the BMN operators, except that the weight  of 
each term is more
complicated than just a phase. Let us illustrate this by explicitly
computing the eigen-operators with 2, 3 and 4 fields.

The entries of the monodromy matrix satisfy recurrence relations
indexed by the number of sites in the chain. These relations follow
from the expression of $T$ in terms of the R-matrix. Denoting by
$A_N$, $B_N$, $C_N$ and $D_N$ the entries of the monodromy matrix for
a chain with $N$ sites, the R-matrix is given by
\begin{equation}
R(\lambda)=P^+\otimes A_1 + P^-\otimes D_1 + \sigma^+\otimes B_1 
+ \sigma^-\otimes C_1 
\end{equation}
with $A_1,~B_1,~C_1$ and $D_1$ given by (\ref{genR}):
\begin{eqnarray}
A_1(\lambda)=f_u(\lambda)P^++g_u(\lambda)P^-
~~~~~~~
&&B_1(\lambda)=h(\lambda)\sigma^-\\
C_1(\lambda)=k(\lambda)\sigma^+
~~~~~~~
&&D_1(\lambda)=g_l(\lambda)P^++f_l(\lambda)P^-
\end{eqnarray}
It turns out that the recurrence relations following from (\ref{T})
are partly diagonal, coupling only $A_N$ with $B_N$ and $C_N$ 
with $D_N$. The relevant ones for our purpose are:
\begin{eqnarray}
A_N&=&A_{N-1}\otimes A_1 + B_{N-1}\otimes C_1\\
B_N&=&A_{N-1}\otimes B_1 + B_{N-1}\otimes D_1~~.
\end{eqnarray}

It is fairly easy to see that, due to the cyclicity of the trace, the
normalized eigen-operators with two and three fields are
independent of the entries of the R-matrix. This is no longer the case
for the operators with four fields which are more complex and
(marginally) retain such a dependence:
\begin{eqnarray}
&&{\cal O}_0^4= |\uparrow\uparrow\uparrow\uparrow\rangle
~~~~~
{\cal O}_1^4= |\uparrow\uparrow\uparrow\downarrow\rangle
~~~~~
{\cal O}_3^4= |\uparrow\downarrow\downarrow\downarrow\rangle
~~~~~
{\cal O}_4^4= |\downarrow\downarrow\downarrow\downarrow\rangle\nonumber\\
&&
{\cal O}_2^4= \frac{1}{{N}}\left[
|\uparrow\uparrow\downarrow\downarrow\rangle
+A
|\uparrow\downarrow\uparrow\downarrow\rangle
+A^2
|\downarrow\uparrow\uparrow\downarrow\rangle
\right]\\
&&{N}^2=(1+A^2)^2+A^2~~~~~{\rm with}~~~~~A =
\frac{g_l(\lambda_2)}{f_u(\lambda_2)} 
\nonumber
\end{eqnarray}
Here $N$ was computed in the planar limit. As stated in the beginning
of this section, the structure of the operator ${\cal O}^4_2$ is similar
to that of a BMN operator with two impurities, except that the phase
weighting each term is replaced by powers of $A$. It is fairly easy to
see that this structure persists for longer chains. 


\section{Discussion\label{end}}

The simple observations made in this note suggest a way to attempt to 
engineer a spin chain once a field theory is given. Generally
speaking, once a bosonic field theory with only 4-point interactions 
is given, the interaction terms provide ``boundary conditions'' for
the quantum Yang-Baxter equation. If a solution obeying the
boundary conditions exists, finding it may still be a difficult
algebraic
problem. The boundary conditions provide, however, an ansatz for part 
(if not all) of the terms in the R-matrix. A useful strategy in finding a
solution might be to interpret (if possible) the theory of interest 
as a deformation of a more symmetric theory. 

The example described above admits a number of interesting
generalizations. At the bosonic level and both in the 2-field and
3-field sectors it would be interesting to relax the holomorphy
condition on the eigen-operators. In the spirit of \S\ref{stupid} and
\S\ref{bsc}, one would attempt to construct a multi-parameter
deformation of the $SO(4)$ and $SO(6)$ spin chains and fix the free
parameters by matching its Hamiltonian with the field theory
Lagrangian expressed in terms of real fields. 

Non-holomorphic operators in Wess-Zumino models have
a simpler description. To see this we first notice that
planar F-term interactions do not switch the positions of holomorphic
and anti-homomorphic fields. Thus, given a set of operators with fixed
ordering of chiral and anti-chiral fields, it seems likely
that the corresponding spin chain is inhomogeneous in the sense that
some sites carry different states than the other ones. Inclusion of
gauge fields seems, unfortunately, to spoil this simple picture, 
since the D-term interactions do not preserve the ordering of holomorphic
and anti-homomorphic fields.


\section*{Acknowledgments}
We would like  to thank Iosif Bena, Nelia Mann, Mark Srednicki,
Mark Spradlin and Anastasia Volovich 
for useful discussions and comments on the manuscript.
This research was supported in part by the National Science Foundation
under Grant No.~PHY00-98395 as well as by the
Department of Energy under Grant No.~DE-FG02-91ER40618.



\end{document}